# Identifying epileptogenic abnormalities through spatial clustering of MEG interictal band power

Thomas W. Owen[1], Vytene Janiukstyte[1], Gerard R. Hall[1], Jonathan J. Horsley[1], Andrew McEvoy[3,5], Anna Miserocchi[3,5], Jane de Tisi[3,4,5], John S. Duncan[3,4,5], Fergus Rugg-Gunn[3,5], Yujiang Wang[1,2,3,5], Peter N. Taylor[1,2,3,5]

1. CNNP Lab (www.cnnp-lab.com), Interdisciplinary Computing and Complex BioSystems Group, School of Computing, Newcastle University, Newcastle upon Tyne, United Kingdom
2. Faculty of Medical Sciences, Newcastle University, Newcastle upon Tyne, United Kingdom
3. UCL Queen Square Institute of Neurology, Queen Square, London, WC1N 3BG, United Kingdom
4. NIHR University College London Hospitals Biomedical Research Centre, UCL Queen Square Institute of Neurology, London WC1N 3BG, United Kingdom
5. National Hospital for Neurology & Neurosurgery, Queen Square, London, WC1N 3BG, United Kingdom

* t.w.owen1@newcastle.ac.uk & peter.taylor@newcastle.ac.uk Orcid ID 0000-0003-2144-9838

## Abstract

Successful epilepsy surgery depends on localising and resecting cerebral abnormalities and networks that generate seizures. Abnormalities, however, may be widely distributed across multiple discontiguous areas. We propose spatially constrained clusters as candidate areas for further investigation, and potential resection. We quantified the spatial overlap between the abnormality cluster and subsequent resection, hypothesising a greater overlap in seizure-free patients.

Thirty-four individuals with refractory focal epilepsy underwent pre-surgical resting-state interictal MEG recording. Fourteen individuals were totally seizure free (ILAE 1) after surgery and 20 continued to have some seizures post-operatively (ILAE 2+). Band power abnormality maps were derived using controls as a baseline. Patient abnormalities were spatially clustered using the k-means algorithm. The tissue within the cluster containing the most abnormal region was compared with the resection volume using the dice score.

The proposed abnormality cluster overlapped with the resection in 71% of ILAE 1 patients. Conversely, an overlap only occurred in 15% of ILAE 2+ patients. This effect discriminated outcome groups well (AUC=0.82).

Our novel approach identifies clusters of spatially similar tissue with high abnormality. This is clinically valuable, providing (i) a data-driven framework to validate current hypotheses of the epileptogenic zone localisation or (ii) to guide further investigation.

# Introduction

Neurosurgery is a treatment option for individuals with refractory focal epilepsy. Surgical success relies on the accurate delineation and resection of the epileptogenic zone (EZ), an area of tissue needed to generate seizures[1]. At present, localization of the EZ typically uses visual analysis of structural and functional imaging data. In recent years, numerous studies have developed quantitative markers of the EZ[2–7]. Both qualitative and quantitative approaches may identify multiple widespread abnormalities in a patient, leading to uncertainty over the design of focal surgery. Thus, in considering abnormalities for further investigation, their spatial proximity should be taken into account.

One approach to develop quantitative markers of the EZ involves neurophysiological abnormality mapping[4,7–9]. This approach posits that regions with abnormal neural dynamics (e.g. measured using band power), relative to normative baselines, should be targeted by surgery to achieve seizure freedom[4,7,9]. A key limitation of such studies is the potential for false-positive abnormalities, which may be far from the EZ, and could lead to mislocalisation and poor outcomes[7].

We developed a framework to account for neurophysiology abnormalities, and their spatial proximity. We use clustering techniques to group areas which are both highly abnormal and spatially similar. We hypothesised a greater overlap between the abnormal cluster and subsequent resection in patients who were seizure free after surgery. Second, we hypothesise that identifying clusters of abnormality has clinical value to assist localization of the EZ.

# Methods

## Patients and abnormality mapping

Resting-state MEG recordings were acquired for thirty-four refractory neocortical patients and seventy healthy controls using a 275 channel whole head CTF MEG scanner. Individual subject

T1-weighted MRI was performed using a 3T GE Signa HDx scanner. Fourteen patients were entirely seizure free (ILAE 1) one year post-operatively[10]. Patient specific maps of eyes-closed interictal band power abnormalities were constructed using healthy data as a baseline. Regional abnormalities in five frequency bands (delta; 1-4Hz, theta; 4-8Hz, alpha; 8-13Hz, beta; 13-30Hz, gamma; 30-47.5Hz and 52.5-80Hz) were estimated and downsampled by taking the maximum absolute z-score across frequencies. A complete description of data acquisition, pre-processing and abnormality mapping has been outlined previously[7]. Subject data and clustering results are summarised in Table 1.

## K-means Clustering

We used k-means clustering to account for the spatial similarity of regions when proposing tissue for resection. K-means is an unsupervised clustering technique that assigns observations into k clusters, minimising the intra-cluster sum of squared distances. Four features were selected to cluster the data. The first three correspond to the standard space x,y, and z coordinates of each neocortical region of interest (ROI) centroid. The fourth feature corresponds to the band power abnormality within each ROI. All four features were mean centred and scaled by the standard deviation to minimise the bias when computing measures of cluster distance. Once partitioned into k clusters, the cluster containing the most abnormal neocortical region was chosen for further investigation. Finally, we constrained the cluster to a single hemisphere by retaining only regions in the most common hemisphere. Regions were constrained to a single hemisphere to better reflect focal resections, which would not span multiple hemispheres. These methods provide a data-driven approach to propose a spatially constrained cluster, which can be considered for resection or further investigation. Note that this approach means that different patients can have different sized clusters, which conforms to clinical practice of different patients with different resection sizes.

In addition to the number of clusters (k), we varied the number of strongest cortical abnormalities (N) used to identify the clusters of abnormality. The optimal parameter set was chosen using a leave-one-out procedure. The parameter pair that maximised the separability of surgical outcome groups in the remaining thirty-three patients was chosen as the optimal for the hold-out patient. The top N abnormalities were chosen over a set z-score threshold as the magnitudes of abnormalities in patients have been shown to relate to duration of epilepsy[9,11] and surgical outcome[7]. Candidate values of k, 2-4, were selected based on visual inspection of patient scree-plots and the N strongest abnormalities were thresholded between 30 and 70 as this range provides a good balance between sensitivity and specificity.

## Validation of the abnormality cluster

The overlap between the abnormality cluster and the resection was quantified using the dice score (DSC). The DSC, defined in equation (1), is the ratio of the overlap between the abnormality cluster and the subsequent resection (true positive; 2TP) relative to the union of the two (2TP+FP+FN). Ranging from zero to one, smaller dice scores correspond to lower overlap between the abnormality cluster and resection, whilst larger dice scores correspond to a higher overlap. Finally, we used the area under the receiver operator curve (AUC) to quantify the separability of surgical outcome groups based on their corresponding dice scores. A one-tailed Mann-Whitney U test was performed to assess the significance of the surgical outcome group separability. The identification of clusters and quantification of the overlap between the abnormality cluster and resection are illustrated in figure 1.

$$\mathrm{DSC} = \frac{2\mathrm{TP}}{2\mathrm{TP}+\mathrm{FP}+\mathrm{FN}} \qquad (1)$$

## Code and data availability

Data and code to reproduce the figures are available upon reasonable request.

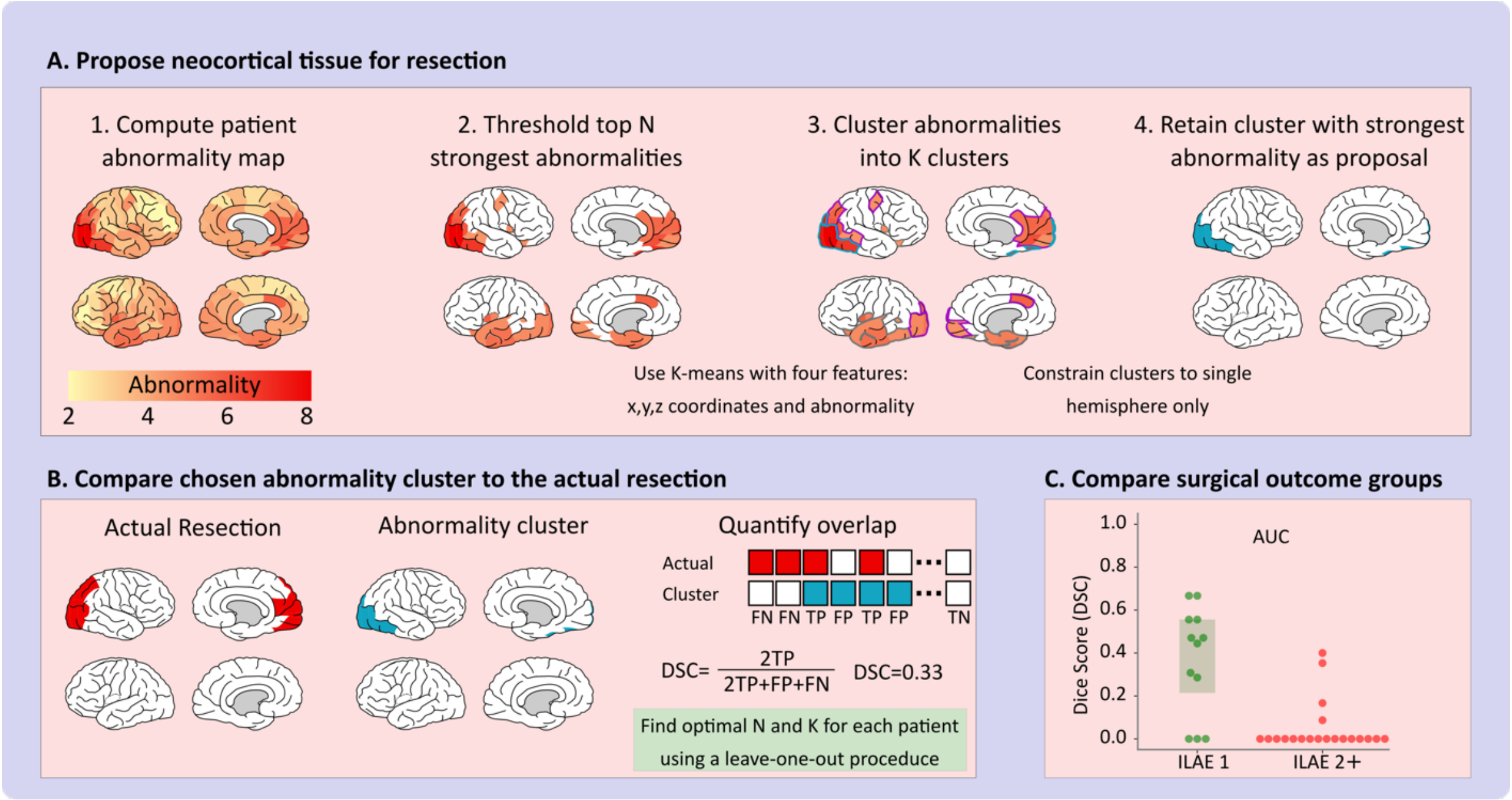

***Figure 1: Processing pipeline to identify abnormality clusters. (A)*** *First, patient abnormality maps were constructed using interictal band power MEG recordings and healthy recordings as a baseline. Second, patient abnormality maps are filtered to retain only the top N strongest abnormalities. Third, the strongest abnormalities are grouped into K clusters using k-means clustering with four features (x,y,z coordinates of each ROI and the abnormality value). Each feature is scaled to minimise any bias in the clustering. Finally, the cluster containing the most abnormal region is selected, with the added constraint that regions reside in a single hemisphere to better reflect focal resections. The chosen hemisphere was decided based on which contained over half of the regions within the cluster.* ***(B)*** *The overlap between the abnormality cluster and actual resection was compared using the dice score (DSC) which measures the ratio of overlap (2TP) relative to union of the abnormality cluster and resection (2TP+FP+FN).* ***TP****: True positive,* ***FP****: False positive,* ***FN****: False negative,* ***TN****: True negative. The DSC varies from 0-1 with DSC= $0$ corresponding to no overlap at all, and DSC= $1$ corresponding to the perfect overlap (identical regions). Steps* ***(A)*** *and* ***(B)*** *are repeated to find the optimal values of N and K to maximise the DSC across the cohort whilst balancing model complexity.* ***(C)*** *The DSC across the whole cohort are compared with differences between seizure-free (ILAE 1) and not-seizure-free (ILAE 2+) quantified using the area under the receiver operator curve (AUC).*

## Results

Using pre-operative interictal MEG band power abnormality maps we sought to propose plausible areas of high abnormality that could contain the EZ, whilst accounting for the spatial proximity of neocortical regions. Two clustering parameters were tuned for each patient using a leave-one-out approach which maximises the separability of surgical outcome groups in the remaining patients. The optimal clustering parameters were identical for all patients and were identified as the top 50 strongest neocortical abnormalities (N), partitioning into three clusters (k) based on the results of the leave-one-out procedure.

Two example patients illustrate the difference between surgical outcome groups (figure 2A,B). For the seizure-free patient there is a high overlap between the abnormality cluster and the subsequent surgical resection in the occipital lobe (figure 2A). This overlap is quantified with a dice similarity of 0.53. Conversely, for the poor outcome patient, there is no overlap between the abnormality cluster and the resection, with the abnormality cluster localised in a different lobe (figure 2 B). These results suggest that our data-driven clustering of abnormalities may provide localising information.

Expanding the analysis to the full cohort (figure 2 C) it is evident that stronger dice scores are attributed with a positive surgical outcome. Indeed, 71% of ILAE 1 patients had some overlap, suggesting even partial resection of our abnormality cluster predisposes a high chance of seizure-freedom. Furthermore, 85% of patients with persistent seizures (ILAE 2+) had zero overlap, suggesting potential mislocalisation in those patients. In addition, this effect successfully discriminated outcome groups (AUC=0.82, p=0.0001). Individuals with overlap between the abnormality cluster and actual resection were fourteen times more likely to be seizure-free post-operatively at 12 months (Odds Ratio: 14.2, 95% confidence interval: [2.6, 76.7]). Similar results were found for other measures of overlap (see supplementary analysis). Moreover, the overlap between the proposed cluster of abnormality and resection was associated with long term seizure-freedom (AUC=0.70, 12 ILAE 1, 22 ILAE 2+). Together, our results suggest that clustering the strongest patient-specific abnormalities may provide clinically useful proposals of tissue that may contain the EZ.

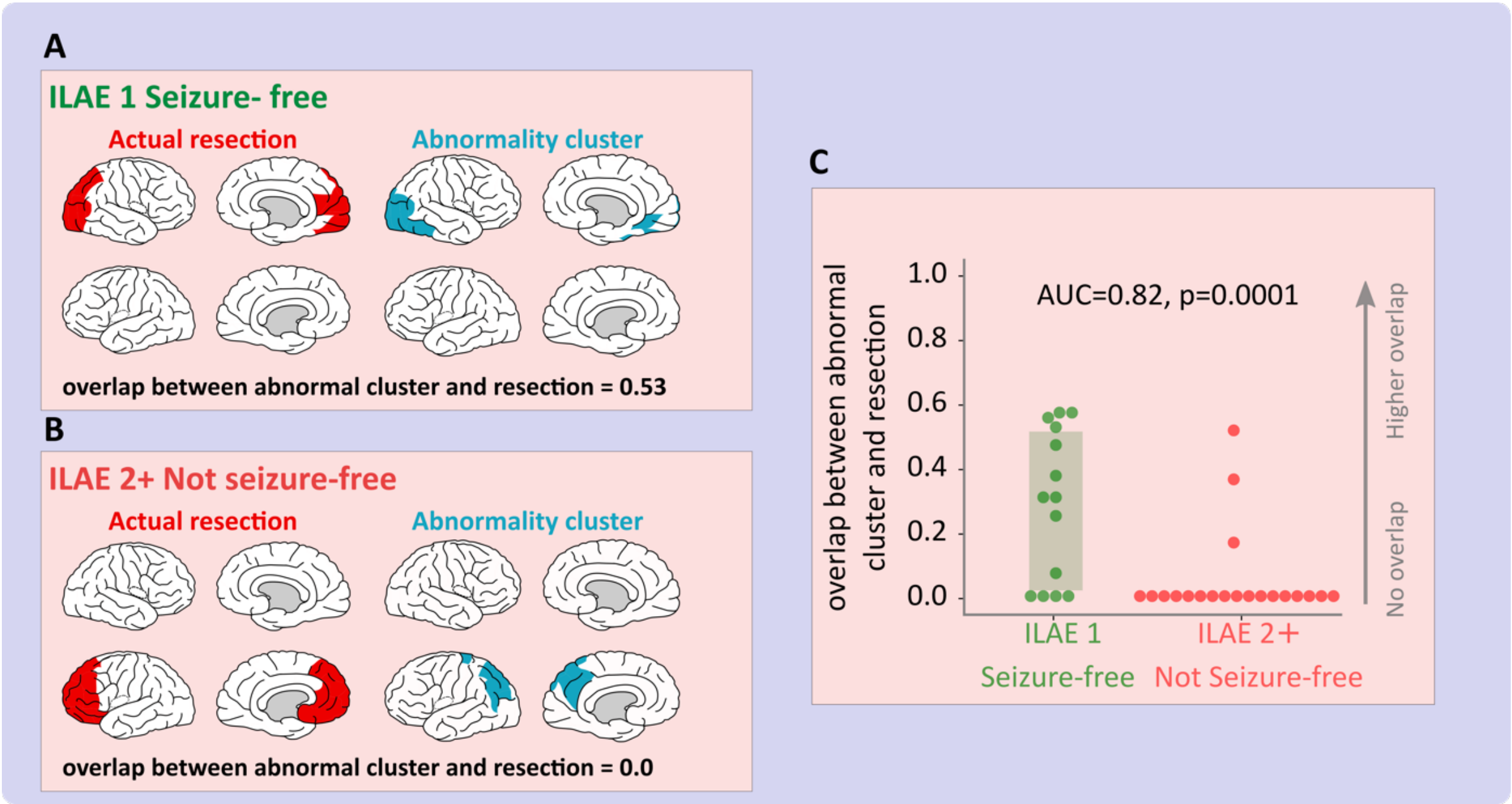

***Figure 2: Overlap between the abnormality cluster and actual resection. (A)*** *Illustration of the overlap between the abnormality cluster and actual resection in an example seizure-free patient. Visually, a high overlap exists between the actual resection (red) and the abnormality cluster (blue). This apparent with a DSC=* $0.53$ *indicating a good overlap. In this scenario we hypothesised a good surgical outcome as both resection masks cover similar tissue. Conversely,* ***(B)*** *corresponds to an example poor outcome patient with no overlap between the abnormality cluster and the resection. A DSC=* $0$ *would suggest a poor outcome for this patient as the abnormality cluster and the resection are in complete disagreement. Cohort-wide comparisons* ***(C)*** *demonstrate that in seizure free patients (green) as hypothesised there is a higher overlap between the abnormality cluster and the subsequent resection in contrast to patients with poor surgical outcomes (red), AUC=0.82, p=0.0001.*

## Discussion

In this study we introduced a novel framework to identify spatially proximal abnormality clusters. We demonstrated that abnormality clusters overlap with the subsequent resection in almost all seizure-free patients. In contrast, most patients who were not seizure-free had resections which were completely discordant with the abnormality cluster. We suggest that abnormality clusters could be clinically useful to assist in the delineation of cortex to be resected or investigated further to cure drug resistant focal epilepsy.

Our current study could complement the existing literature that develop markers of the EZ using structural[2,12,13] and functional[4,7,14–16] data. The clinical adoption of previous work is difficult since markers of the EZ may neglect spatial properties of the tissue. As such, proposals of which tissue to resect may not be spatially contiguous, with abnormalities spanning multiple lobes or hemispheres. The benefits of accounting for spatial properties are well established in neuroimaging using, e.g. random field theory[17]. Random field theory has been applied in the context of epilepsy[18], with studies identifying clusters of statistically abnormal voxels in patients with epileptogenic malformations such as focal cortical dysplasia. In the context of surgical resections we incorporated the spatial proximity of cortical abnormalities using k-means clustering, demonstrating that simple clustering techniques could be applied to current markers of epileptogenic tissue in order to suggest realistic data-driven areas of tissue for resection.

It has been established that the resection of epileptogenic tissue is associated with seizure freedom post-surgically[1]. A stronger overlap between the abnormality cluster and resection in seizure-free patients suggests that our abnormality clusters may indeed be capturing the EZ. Interestingly, even in patients rendered seizure-free, the overlap is not perfect (Figure 2). The discrepancies between the abnormality cluster and resection could be attributed to a number of factors. First, targeting the epileptogenic tissue for resection may require that healthy tissue is also removed, tissue which is not currently considered within our framework. As such, this would increase the false negative rate, thus decreasing the dice overlap. Alternatively, a decrease in the overlap could be attributed to a difficult to localise EZ, either due to inconsistencies across modalities or an extensive epileptogenic network [9]. As a result, a larger area of tissue may be targeted to ensure the total resection of the EZ. This may have been the case in Figure 2A with the resection including the inferior and superior parietal lobe. Finally, the suggested abnormality cluster may contain abnormal tissue that visually appears normal in neuroimaging and neurophysiology modalities such as MRI and MEG. For example, the abnormality cluster in Figure 2A extends to the inferior temporal lobe, areas of tissue which was not targeted in during the resection. The results of this study suggest that the abnormality cluster contains complementary information which could aid in the accurate laterlaisation and localisation of the EZ.

One limitation of our current clustering framework is the added constraint that abnormalities should reside in a single hemisphere. Although the constraint of single hemisphere clusters better reflects a focal resection, it is conceivable that the strongest abnormalities could span both hemispheres. As such by constraining abnormalities to a single hemisphere we are discarding potentially valuable information that may alter clinical decision making. Future studies could investigate whether intracranial EEG implantation is associated with MEG band power abnormalities and determine if the cluster of strongest abnormalities provide complementary information to guide electrode implantation in seemingly healthy tissue across both hemispheres.

The accurate delineation and resection of the EZ has been shown to be associated with surgical success. We introduce a fully data-driven clustering technique to spatially constrain markers of the EZ into plausible clusters of abnormality. Our approach could be clinically valuable, offering a data-driven cluster of tissue believed to contain the EZ. This can be used for validation against the current candidate tissue for resection, or to guide further investigation.

## Acknowledgements

We thank members of the Computational Neurology, Neuroscience & Psychiatry Lab ( www.cnnp-lab.com ) for discussions on the analysis and manuscript. The normative data collection was supported by an MRC UK MEG Partnership Grant, MR/K005464/1. T.O. and J.J.H were supported by the Centre for Doctoral Training in Cloud Computing for Big Data (EP/L015358/1). P.N.T. and Y.W. are both supported by UKRI Future Leaders Fellowships (MR/T04294X/1, MR/V026569/1). J.S.D is supported by the Wellcome Trust Innovation grant 218380. J.S.D and J.dT are supported by the NIHR University College London Hospitals Biomedical Research Centre, UCL Queen Square Institute of Neurology, London WC1N 3BG, United Kingdom.

## Author contributions

Thomas W. Owen, Yujiang Wang, and Peter N. Taylor contributed to the conception and design of the study. Thomas W. Owen, Vytene Janiukstyte, Gerard R. Hall, Jonathan J. Horsley, Andrew McEvoy, Anna Miserocchi, Jane de Tisi, John S. Duncan, Fergus Rugg-Gunn, Yujiang Wang, and Peter N. Taylor, contributed to the acquisition and analysis of data. Thomas W. Owen, and Peter N. Taylor contributed to drafting of the text and preparing the figures.

## Conflict of interest

No relevant conflicts of interest are reported.

# Supplementary

## Analysis using other measures of overlap

One potential drawback of the dice score is that it penalises the degree of overlap based on both the false positive and false negative rate. That is, the degree of overlap is dampened if the actual resection extends beyond the abnormality cluster, or if the abnormality cluster extends to areas outside of the actual resection. We assess the degree of overlap between the abnormality cluster using alternative measures, namely the positive predictive value (PPV). The PPV measure neglects the effects of the false negative rate, capturing the ratio of the overlap (TP) relative to the total number of regions in either the abnormality cluster, or actual resection. We compare the degree of overlap between surgical outcome groups using (1) the PPV that quantifies the proportion of the abnormality cluster that intersects with the actual resection relative to all regions within the abnormality cluster, and (2) the PPV measuring the proportion of the actual resection within the abnormality cluster. Parameter values of K=3 and N=50 were used for all patients to directly compare with the dice score in main text (Figure 2). The results of the PPV measures studies are illustrated in Figure 3. Comparable results to the dice score can be seen for both measures of the PPV. Similar AUC values to the dice score indicate that neither the actual resection, nor abnormality cluster extending into alternative regions have a negative impact on the separability of outcome groups. However, each measure provides a different perspective on the overlap, providing clinical teams with additional clarity and information during pre-surgical planning.

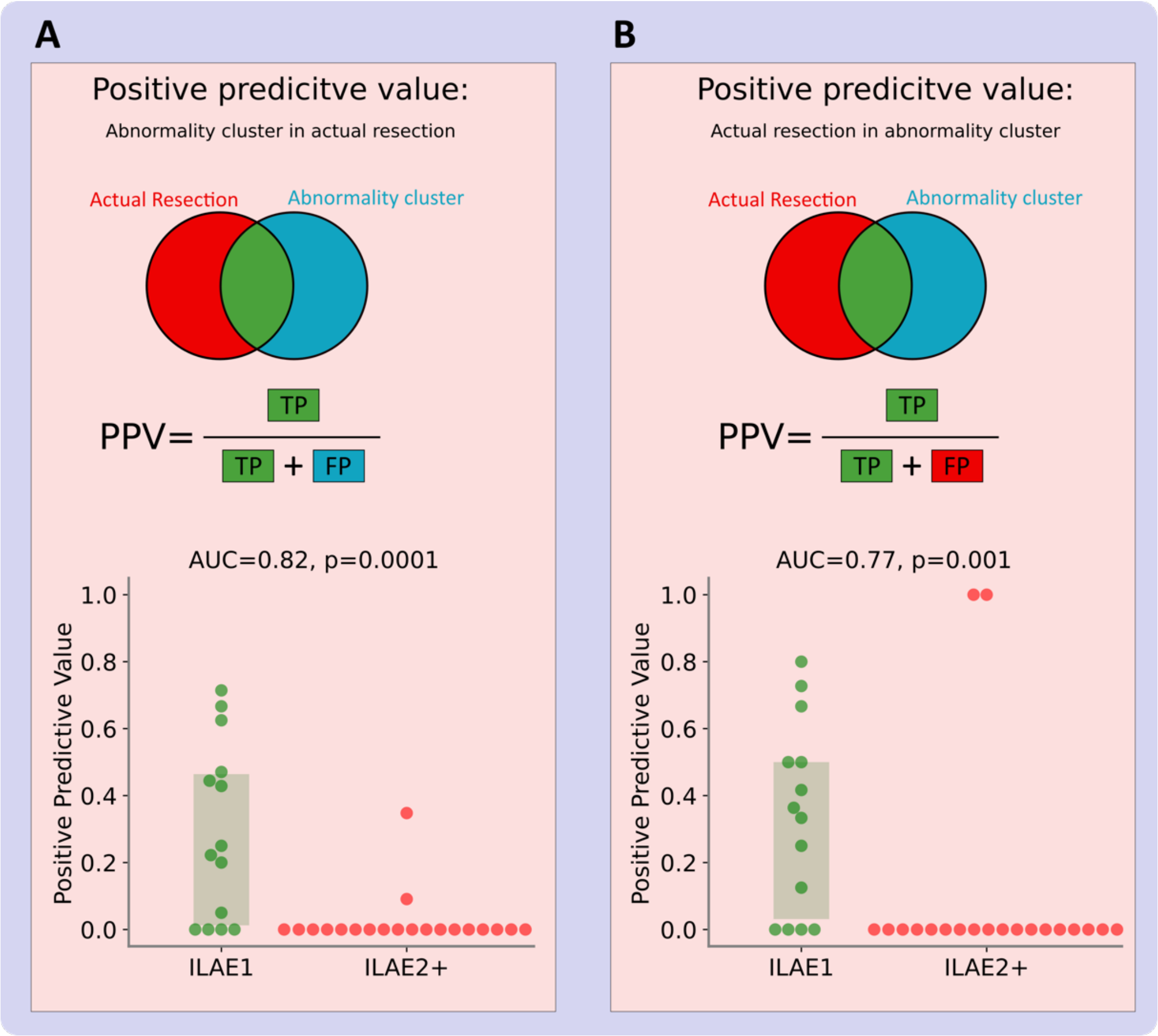

***Overlap between the abnormality cluster and actual resection using the positive predictive values.*** *The degree over overlap between the abnormality cluster and actual resection were compared across surgical outcome groups using the positive predictive value (PPV).* ***(A)*** *First we quantify how much of the abnormality cluster overlaps with the actual resection, thus accounting for proposed regions that extend beyond the true resection. The degree of overlap differed significantly between surgical outcome groups (AUC=0.82, p=0.0001).* ***(B)*** *The second PPV measure captures the proportion of the actual resection that overlaps with the abnormality cluster. This measure also accounts for the regions that were actually resected but not captured within the abnormality cluster. A significant separability of outcome groups was also detected (AUC=0.77, p=0.001) indicating that both measures could be leveraged to aid clinical decision making.*

## Table of patient data

***Summary of patient metadata and clustering results.*** *Commonly acquired patient metadata are reported including the side of surgical resection, localisation of the resection (F: Frontal, T: Temporal, P: Parietal, O: Occipital), one year post-operative outcome, long term surgical outcome, and whether the patient relapsed at a later date. Additionally, the dice score, measuring the overlap between the abnormality cluster and actual resection is reported.*

| Patient ID | Side | Resection Site | Surgical Outcome (1 year) | Surgical Outcome (Long term) | Years of follow up | Relapsed (Yes/No) | Dice Score |
|---|---|---|---|---|---|---|---|
| 1 | L | F | ILAE 2+ | ILAE 2+ | 5 | - | 0 |
| 2 | L | F | ILAE 2+ | ILAE 2+ | 5 | - | 0 |
| 3 | R | F | ILAE 2+ | ILAE 2+ | 5 | - | 0 |
| 4 | L | F | ILAE 1 | ILAE 1 | 2 | No | 0.308 |
| 5 | L | F | ILAE 1 | ILAE 2+ | 5 | Yes | 0.375 |
| 6 | R | F | ILAE 2+ | ILAE 1 | 5 | - | 0 |
| 7 | R | F | ILAE 2+ | ILAE 1 | 5 | - | 0.516 |
| 8 | L | F | ILAE 2+ | ILAE 2+ | 5 | - | 0 |
| 9 | R | P | ILAE 1 | ILAE 1 | 5 | No | 0 |
| 10 | L | F | ILAE 1 | ILAE 1 | 5 | No | 0.571 |
| 11 | L | F | ILAE 2+ | ILAE 2+ | 3 | - | 0 |
| 12 | R | T | ILAE 1 | ILAE 1 | 5 | No | 0 |
| 13 | L | F | ILAE 2+ | ILAE 2+ | 4 | - | 0 |
| 14 | R | P | ILAE 2+ | ILAE 1 | 5 | - | 0.364 |
| 15 | L | O | ILAE 1 | ILAE 2+ | 5 | Yes | 0 |
| 16 | R | F | ILAE 2+ | ILAE 2+ | 2 | - | 0 |
| 17 | R | F | ILAE 1 | ILAE 2+ | 5 | Yes | 0.556 |
| 18 | R | P | ILAE 2+ | ILAE 2+ | 5 | - | 0 |
| 19 | L | T | ILAE 1 | ILAE 1 | 5 | No | 0 |
| 20 | L | T | ILAE 2+ | ILAE 2+ | 5 | - | 0 |
| 21 | L | T | ILAE 2+ | ILAE 2+ | 5 | - | 0 |
| 22 | L | F | ILAE 1 | ILAE 1 | 5 | No | 0.308 |
| 23 | L | FP | ILAE 2+ | ILAE 2+ | 5 | - | 0.167 |
| 24 | R | F | ILAE 2+ | ILAE 2+ | 5 | - | 0 |
| 25 | L | F | ILAE 1 | ILAE 1 | 5 | No | 0.471 |
| 26 | L | F | ILAE 2+ | ILAE 2+ | 4 | - | 0 |
| 27 | L | F | ILAE 2+ | ILAE 2+ | 5 | - | 0 |
| 28 | R | OP | ILAE 1 | ILAE 1 | 5 | No | 0.25 |

| Patient ID | Side | Resection Site | Surgical Outcome (1 year) | Surgical Outcome (Long term) | Years of follow up | Relapsed (Yes/No) | Dice Score |
|---|---|---|---|---|---|---|---|
| 29 | R | F | ILAE 1 | ILAE 1 | 5 | No | 0.571 |
| 30 | R | O | ILAE 1 | ILAE 2+ | 5 | Yes | 0.526 |
| 31 | L | OP | ILAE 2+ | ILAE 2+ | 5 | - | 0 |
| 32 | R | F | ILAE 2+ | ILAE 2+ | 4 | - | 0 |
| 33 | R | P | ILAE 2+ | ILAE 2+ | 5 | - | 0 |
| 34 | R | F | ILAE 1 | ILAE 2+ | 5 | Yes | 0.071 |

## References

1. Rosenow F, Lüders H. Presurgical evaluation of epilepsy. Brain, 2001 124(9):1683–700. Available from: https://doi.org/10.1093/brain/124.9.1683

2. Sinha N, Wang Y, Silva NM da, Miserocchi A, McEvoy AW, Tisi J de, et al. Node abnormality predicts seizure outcome and relates to long-term relapse after epilepsy surgery, Neuroscience; 2019 [cited 2020]. Available from: http://biorxiv.org/lookup/doi/10.1101/747725

3. Papadelis C, Perry MS. Localizing the Epileptogenic Zone with Novel Biomarkers. Seminars in Pediatric Neurology, 2021 39:100919. Available from: https://www.sciencedirect.com/science/article/pii/S1071909121000474

4. Taylor PN, Papasavvas CA, Owen TW, Schroeder GM, Hutchings FE, Chowdhury FA, et al. Normative brain mapping of interictal intracranial EEG to localize epileptogenic tissue. Brain, 2022 :awab380. Available from: https://doi.org/10.1093/brain/awab380

5. Morgan VL, Sainburg LE, Johnson GW, Janson A, Levine KK, Rogers BP, et al. Presurgical temporal lobe epilepsy connectome fingerprint for seizure outcome prediction. Brain Communications, 2022 :fcac128. Available from: https://doi.org/10.1093/braincomms/fcac128

6. Corona L, Tamilia E, Perry MS, Madsen JR, Bolton J, Stone SSD, et al. Non-invasive mapping of epileptogenic networks predicts surgical outcome. Brain, 2023 :awac477. Available from: https://doi.org/10.1093/brain/awac477

7. Owen TW, Schroeder GM, Janiukstyte V, Hall GR, McEvoy A, Miserocchi A, et al. MEG abnormalities and mechanisms of surgical failure in neocortical epilepsy. Epilepsia, 2023 n/a(n/a). Available from: https://doi.org/10.1111/epi.17503

8. Frauscher B, Ellenrieder N von, Zelmann R, Doležalová I, Minotti L, Olivier A, et al. Atlas of the normal intracranial electroencephalogram: Neurophysiological awake activity in different cortical areas. Brain: A Journal of Neurology. 2018; 141(4):1130–44.

9. Bernabei JM, Sinha N, Arnold TC, Conrad E, Ong I, Pattnaik AR, et al. Normative intracranial EEG maps epileptogenic tissues in focal epilepsy. Brain, 2022 :awab480. Available from: https://doi.org/10.1093/brain/awab480

10. Wieser HG, Blume WT, Fish D, Goldensohn E, Hufnagel A, King D, et al. Proposal for a New Classification of Outcome with Respect to Epileptic Seizures Following Epilepsy Surgery. Epilepsia, 2001 42(2):282–6. Available from: https://onlinelibrary.wiley.com/doi/abs/10.1046/j.1528-1157.2001.35100.x

11. Owen TW, Tisi J de, Vos SB, Winston GP, Duncan JS, Wang Y, et al. Multivariate white matter alterations are associated with epilepsy duration. European Journal of Neuroscience, 2021 53(8):2788–803. Available from: https://onlinelibrary.wiley.com/doi/abs/10.1111/ejn.15055

12. Bonilha L, Jensen JH, Baker N, Breedlove J, Nesland T, Lin JJ, et al. The brain connectome as a personalized biomarker of seizure outcomes after temporal lobectomy. Neurology. 2015; 84(18):1846–53.

13. Keller SS, Glenn GR, Weber B, Kreilkamp BAK, Jensen JH, Helpern JA, et al. Preoperative automated fibre quantification predicts postoperative seizure outcome in temporal lobe epilepsy. Brain, 2017 140(1):68–82. Available from: https://academic.oup.com/brain/article/140/1/68/2670176

14. Englot DJ, Hinkley LB, Kort NS, Imber BS, Mizuiri D, Honma SM, et al. Global and regional functional connectivity maps of neural oscillations in focal epilepsy. Brain, 2015 138(8):2249–62. Available from: https://www.ncbi.nlm.nih.gov/pmc/articles/PMC4840946/

15. Nissen IA, Stam CJ, Reijneveld JC, Straaten IECW van, Hendriks EJ, Baayen JC, et al. Identifying the epileptogenic zone in interictal resting-state MEG source-space networks. Epilepsia, 2017 58(1):137–48. Available from: https://onlinelibrary.wiley.com/doi/abs/10.1111/epi.13622

16. Ramaraju S, Wang Y, Sinha N, McEvoy AW, Miserocchi A, Tisi J de, et al. Removal of Interictal MEG-Derived Network Hubs Is Associated With Postoperative Seizure Freedom. Frontiers in Neurology, 2020 11. Available from: https://www.frontiersin.org/articles/10.3389/fneur.2020.563847/full

17. Frackowiak RSJ, Friston KJ, Frith CD, Dolan RJ, Price CJ, Zeki S, et al., editors. Chapter 44 - Introduction to Random Field Theory. In: Human Brain Function (Second Edition), Burlington: Academic Press; 2004 [cited 2023]. p. 867–79. Available from: https://www.sciencedirect.com/science/article/pii/B9780122648410500469

18. Hong S-J, Bernhardt BC, Schrader DS, Bernasconi N, Bernasconi A. Whole-brain MRI phenotyping in dysplasia-related frontal lobe epilepsy. Neurology, 2016 86(7):643–50. Available from: https://www.neurology.org/lookup/doi/10.1212/WNL.0000000000002374